\documentclass[%
 aip,
 jmp,%
 amsmath,amssymb,
 preprint,%
]{revtex4-1}

\usepackage{graphicx}% Include figure files
\usepackage{dcolumn}% Align table columns on decimal point
\usepackage{bm}% bold math

\begin{document}

%\preprint{AIP/123-QED}

\title{Effect of realistic metal electronic structure on the lower limit of contact resistivity of epitaxial metal-semiconductor contacts}% Force line breaks with \\

\author{Ganesh Hegde}
	\email{ganesh.h@ssi.samsung.com}
\author{R. Chris Bowen}%
\affiliation{Advanced Logic Lab, Samsung Semiconductor R\&D Center, Austin, TX 78754}
\date{\today}% It is always \today, today,
\begin{abstract}
The effect of realistic metal electronic structure on the lower limit of resistivity in [100] oriented $n$-Si is investigated using full band Density Functional Theory and Semi-Empirical Tight Binding (TB) calculations. Using simulation unit cells guided by the interface chemistry of epitaxial CoSi$_{2}$ on [100] oriented Si observed experimentally, it is shown that the 'ideal metal' assumption fails in some situations and consequently underestimates the lower limit of contact resistivity in $n$-Si by at least an order of magnitude at high doping concentrations. The mismatch in transverse momentum space in the metal and the semiconductor, the so-called 'valley filtering effect', is shown to be dependent on the interface chemistry simulated. The results emphasize the need for explicit inclusion of the metal atomic and electronic structure in the atomistic modeling of transport across metal-semiconductor contacts.
\end{abstract}

\maketitle
%\section{Introduction}
Metal-Semiconductor (M-S) specific contact resistivity ($\rho_c$) is a key metric in the performance of nano-scaled semiconductor device technology. For example, to meet scaling requirements, the International Technology Roadmap for Semiconductors (ITRS)\cite{itrs} has called for a $\rho_c$ value of $10^{-9} \Omega$-cm$^{2}$ by 2023. 

Much of the existing experimental and theoretical work on improving M-S contact resistivity is driven by semi-classical models of electronic transport through the M-S interface. These models use empirically derived quantities such as Schottky Barrier Height (SBH), width and doping concentrations (see, for instance, the Wentzel, Kramers, Brillouin (WKB)-based model of Yu \cite{yu1970electron}) and have a limited applicability. The WKB approximation, for instance, is valid only in the limit that the barrier potential varies slowly compared to the electron wavelength \cite{griffiths2013introduction}. 

From an atomistic modeling standpoint, Maassen et al. \cite{maassen2013full} recently reported the lower limits of contact resistivity in Si using full band Tight Binding (TB) calculations. They report specific contact resistivities in the low $10^{-11} \Omega$-cm$^{2}$ range at doping concentrations close to the solubility limit of P in Si ($N_d$ $\approx$ 7$\times$10$^{20} cm^{-3}$) \cite{trumbore1960solid}. Weber \cite{weber2013importance} and Park et al.\cite{park2013scaling} have performed Non-Equilibrium Greens Function (NEGF) transport calculations on the M-S interface system that incorporates metallic electronic structure in an effective-mass based approach. While these investigations provide important insights into the transport mechanism at M-S interfaces at a level that surpasses semi-classical approaches, they ignore a number of important effects that manifest themselves at M-S interfaces as device dimensions are scaled down. 

For instance, it is not possible to capture complicated Fermi Surfaces of metal silicides such as NiSi, TiSi and CoSi$_{2}$ accurately in an effective-mass based approach. At extremely high doping concentrations (in the high 10$^{20}$ cm$^{-3}$ range and above), upper valleys in Si begin to conduct, so that the use of a single-band effective mass for Si is no longer valid. It is also unclear from these investigations if the 'ideal metal' approximation of a spherical Fermi Surface (FS) \cite{weber2013importance} with sufficient modes to supply all semiconducting modes \cite{maassen2013full} is realistic. 

Gao et al. \cite{gao2012ab} have investigated the electronic structure of the metal silicide-silicon interface using \textit{ab}-\textit{initio} Density Functional Theory (DFT) based calculations. Due to explicit inclusion of the metal electronic structure in their work, a more realistic picture of the factors governing electronic transport at these interfaces emerges. Their work, however, does not explicitly quantify the effect of including realistic metal electronic structure on eventual contact resistivity or on its lower bound in Si. It is also unclear to what extent epitaxial interface chemistry affects electronic transmission through the interface. 
%One such effect is the 'filtering' of semiconducting valleys by metals due to a mismatch between their transverse electronic structure \cite{weber2013importance,gao2012ab}. Semiconductor valley filtering reduces the number of semiconducting modes available for conduction. Consequently, the contact resistivity is estimated to increase up to an order of magnitude.
%While Gao et al \cite{gao2012ab} only address valley filtering tangentially in an otherwise \textit{ab}-\textit{initio} Density Functional Theory (DFT) investigation of transport at a silicide/silicon interface, Weber\cite{weber2013importance} has addressed the issue directly using the Non-Equlibrium Greens Function (NEGF) transport formalism \cite{datta1997electronic} in the single-sub-band effective-mass approximation.
%The effective mass-NEGF formalism is useful to obtain a quick estimate of how a variety of factors - SBH, image force induced barrier lowering, doping concentration - affect eventual contact resistivity. 
%The phenomenon of semiconductor valley filtering by metal transverse momentum, however, contains a number of subtleties that cannot be captured simply in an effective-mass based approach.

%Additionally, the specific interface chemistry dictated by choice of simulation unit cell results in Fermi Surfaces in the metal and semiconductor that may be far from their bulk, face-centered-cubic (FCC) primitive unit cell counterparts. This might give rise to significantly different valley filtering behavior for the same choice of metal and semiconductor. 
In this letter, the effect of realistic metal electronic structure on the lower limit of contact resistivity in [100] oriented $n$-Si is quantified using full band DFT and TB calculations. Using simulation unit cells guided by the interface chemistry of epitaxial CoSi$_{2}$ on [100] oriented Si observed experimentally, it is shown that the 'ideal metal' assumption fails in some situations and consequently underestimates the lower limit of contact resistivity in $n$-Si by at least an order of magnitude at high doping concentrations. The mismatch in transverse momentum space in the metal and the semiconductor, the so-called 'valley filtering effect' \cite{gao2012ab,weber2013importance}, is shown to be sensitive to the interface chemistry simulated. Finally, the limits to the applicability of such an analysis are also discussed in detail.

\begin{figure*}
\centering
\includegraphics[width=0.8\textwidth]{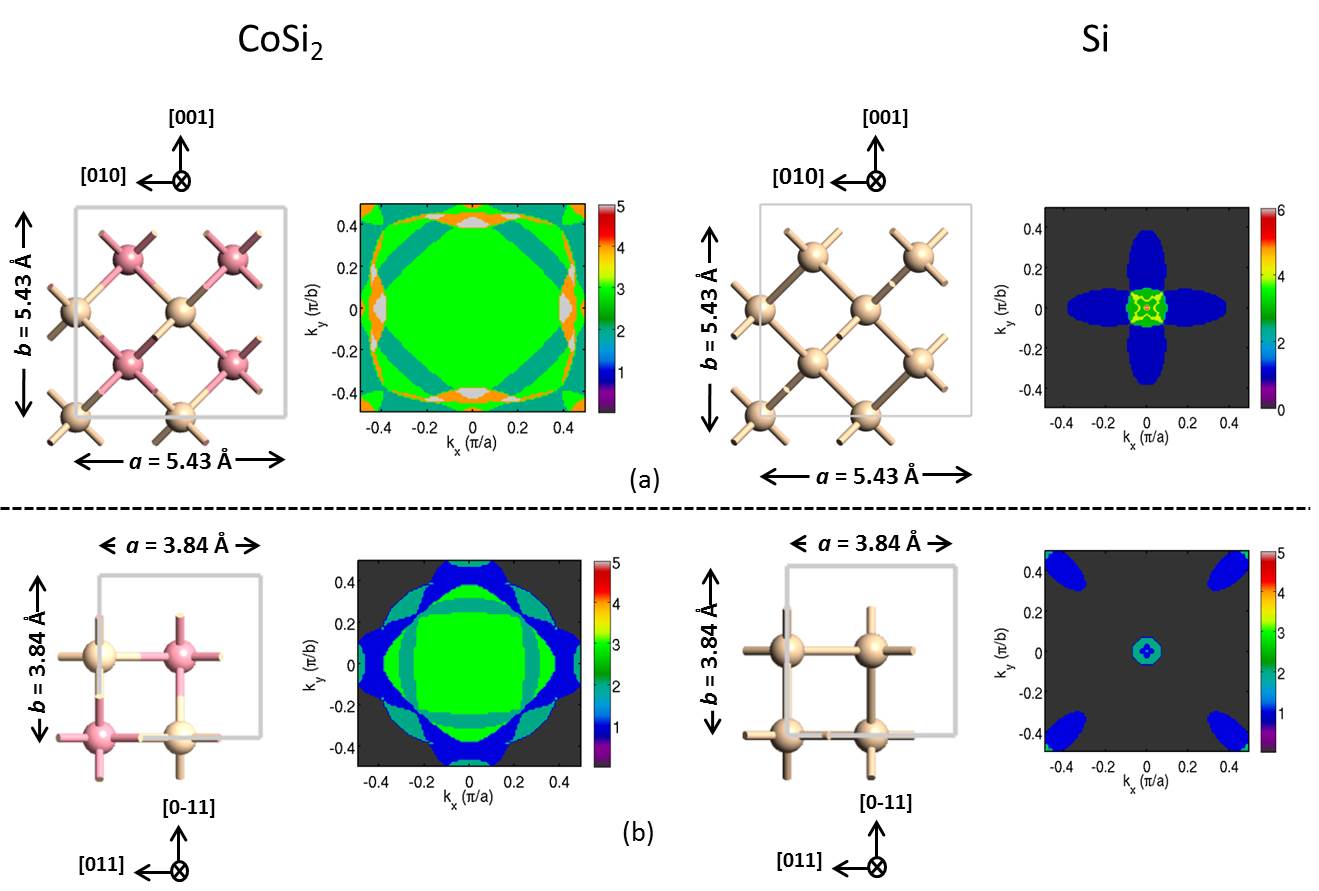}
\caption{Cross section structures and transverse momentum-resolved transmission spectra for epitaxial CoSi$_{2}$ on [100] oriented Si. Part (a) is for the case of [100] oriented CoSi$_{2}$ unit cell bound by the [010] and [001] directions lattice matched to a [100] oriented Si unit cell bound by the same orientations. Part (b) is for the case of [100] oriented CoSi$_{2}$ unit cell bound by the [0-11] and [011] directions lattice matched to a [100] oriented Si unit cell bound by the same orientations. The transmission spectra are obtained at the Fermi Level for the metal and the semiconductor. The doping in Si is $N_d=$ 7$\times$10$^{20} cm^{-3}$, the approximate solid solubility limit of P in Si at annealing temperatures of over 950C \cite{}. Lightly colored spheres represent Si while dark spheres represent Co atoms.} 
\label{fig:Fig1}
\end{figure*}

%\section{Computational Methodology}
The CoSi$_{2}$-Si epitaxial interface system observed experimentally \cite{tung1989homoepitaxial} is chosen as a representative M-S system in this letter. Bulk CoSi$_{2}$ has a 3 atom FCC unit cell with a Co atom at (0,0,0) and Si atoms at ($\frac{a_0}{4}$,$\frac{a_0}{4}$,$\frac{a_0}{4}$) and ($\frac{3a_0}{4}$,$\frac{3a_0}{4}$,$\frac{3a_0}{4}$), where the lattice parameter $a_0=$  5.365$\AA$ at room temperature. The experimental growth of epitaxial CoSi$_{2}$ on [100] oriented $n$-Si at its room temperature lattice parameter of 5.43$\AA$ results in two dominant [100] epitaxial orientations with a mean lattice mismatch of about 0.81\% in CoSi$_{2}$ in the plane transverse to the interface. These are - [100] oriented CoSi$_{2}$ lattice matched to [100] oriented Si, where the substrate is bound by $a.$ [010] and [001] crystal orientations and $b.$ [011] and [0-11] orientations. This motivates the use of two distinct unit cell cross sections for the M-S interface system. These unit cell cross sections are shown in Fig \ref{fig:Fig1}. 
Using these unit cells as the starting point, the transverse-momentum resolved transmission spectra in [100] oriented bulk CoSi$_{2}$ and Si are computed \textit{separately} for a range of energies in the NEGF formalism assuming periodic boundary conditions transverse to the interface growth direction. For the case of ballistic transmission, these transmission spectra represent the projection of the constant energy surface onto the Transverse Brillouin Zone (TBZ). In other words, they represent the number of conducting modes in the direction of transport for each discrete ($k_x$,$k_y$) value. The transmission spectrum of Si is computed using the $sp3d5s^*$ TB parameters of Boykin et al. \cite{boykin2004valence} The Generalized Gradient Approximation (GGA) functional of Perdew, Burke and Ernzerhof (PBE) \cite{perdew1996generalized} is used to compute the transmission of CoSi$_{2}$ in DFT. The Atomistix Tool Kit (ATK) \cite{atk} is used for both sets of calculations. An energy grid of 1 meV was used in all transmission calculations. The TBZ given by $\frac{-\pi}{a}\leq k_x\leq \frac{\pi}{a}$ and $\frac{-\pi}{b}\leq k_y\leq \frac{\pi}{b}$ (where a and b are unit cell lengths transverse to the interface orientation) in the metal and semiconductor for case $a$ was sampled using a uniform $k$-space grid of 100$^2$ $k$ points for case $a$ and a grid of 150$^2$ $k$ points for case $b$. Figure \ref{fig:Fig1} shows the transmission spectra obtained for CoSi$_{2}$ at Si at their respective Fermi levels. The Fermi level in Si corresponds to a doping concentration of $N_d=$ 7$\times$10$^{20} cm^{-3}$.

Once the transmission spectra are computed for the metal and semiconductor separately, the two are 'coupled' together using some simplifying assumptions. First, the potential barrier at the epitaxial interface is neglected. Second, the metal modes at a given energy are assumed to couple perfectly to available semiconducting modes at the same energy without reflections. 
These assumptions imply that transmission across the interface is assumed to be ballistic and every propagating metal mode that finds a corresponding semiconducting mode propagates without attenuation at the interface. Finally, elastic scattering between different transverse momenta in the semiconductor is neglected. In the metal, level broadening due to electron-electron scattering is approximated through a relatively large optical potential $\eta=$ 1 meV which results in an electron-electron scattering rate of the order of 10$^{-13}$ seconds. 

\begin{figure*}
\centering
\includegraphics[width=0.8\textwidth]{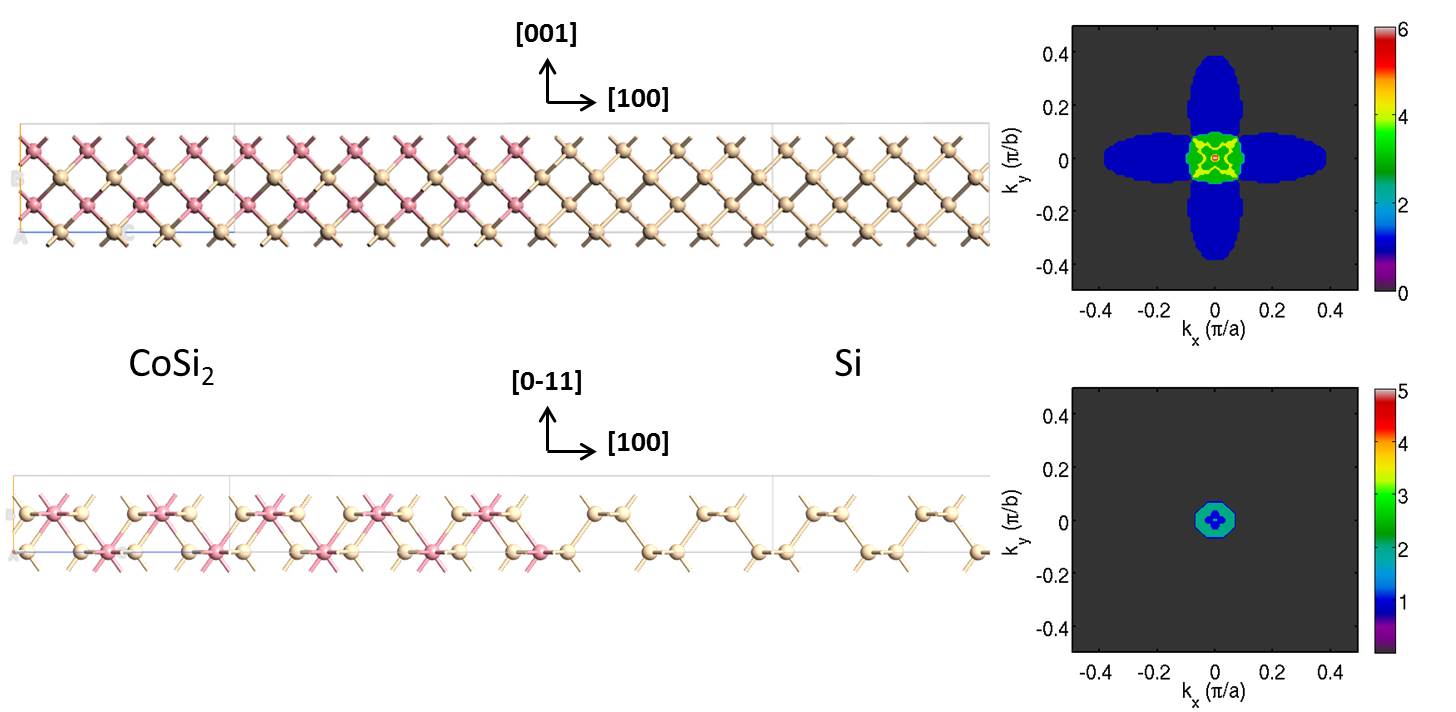}
\caption{Atomic structure and transverse momentum-resolved transmission spectra for the M-S interfaces formed by 'coupling' CoSi$_{2}$ and Si unit cells under the assumptions described above. Due to favorable overlapping of transverse momentum space, it can be seen that the epitaxial interface bounded by the [010] and [001] orientations leaves the transmission spectrum in Si unchanged from its own bulk limit. The interface bound by [0-11] and [011] orientations, however, results in the complete filtering out four out of six Si valleys due to poor overlap between the transverse momentum space of CoSi$_{2}$ and Si for this unit cell orientation as shown in Figure \ref{fig:Fig1}. }
\label{fig:Fig2} 
\end{figure*}

%From a physical perspective, these assumptions are not unrealistic. 
The assumptions made above ensure that a given transverse mode in Si (i.e. an allowed ($k_x, k_y$) value) conducts only if it is supplied by a conducting mode at the same transverse mode in CoSi$_{2}$. These simplifying assumptions facilitate the study of electron transport across the CoSi$_{2}$-Si interface in the limit of ballistic transmission while retaining an accurate representation of metal electronic structure. They also allow for a consistent calculation of the lower limits of specific contact resistivity, regardless of whether the metal is assumed to be ideal or not. Figure \ref{fig:Fig2} shows the transmission spectra that arise as a result of such a coupling for the two interface orientations considered in this work.

Once the transverse momentum-resolved transmission spectra for the coupled interface are computed under the assumptions outlined above, the number of available conducting modes per unit cross sectional area as a function of energy $E$ in Si, $M(E)$, is simply calculated as per the mode-counting formalism outlined in Maassen et al. \cite{maassen2013full}.
\begin{equation}
\label{eq:Eq1}
M(E) = \frac{1}{(2\pi)^{2}}\int_{BZ}M(E,k_x,k_y)dk_xdk_y  \quad (m^{-2})
\end{equation}
%Fig \ref{fig:Fig3}(a) plots this quantity as a function of the offset of the Fermi Level and the conduction band minimum (CBM) in the $n$-Si unit cells coupled to their respective lattice matched CoSi$_{2}$ unit cells shown in Fig \ref{fig:Fig1}.
\begin{figure*}
\centering
\includegraphics[width=0.8\textwidth]{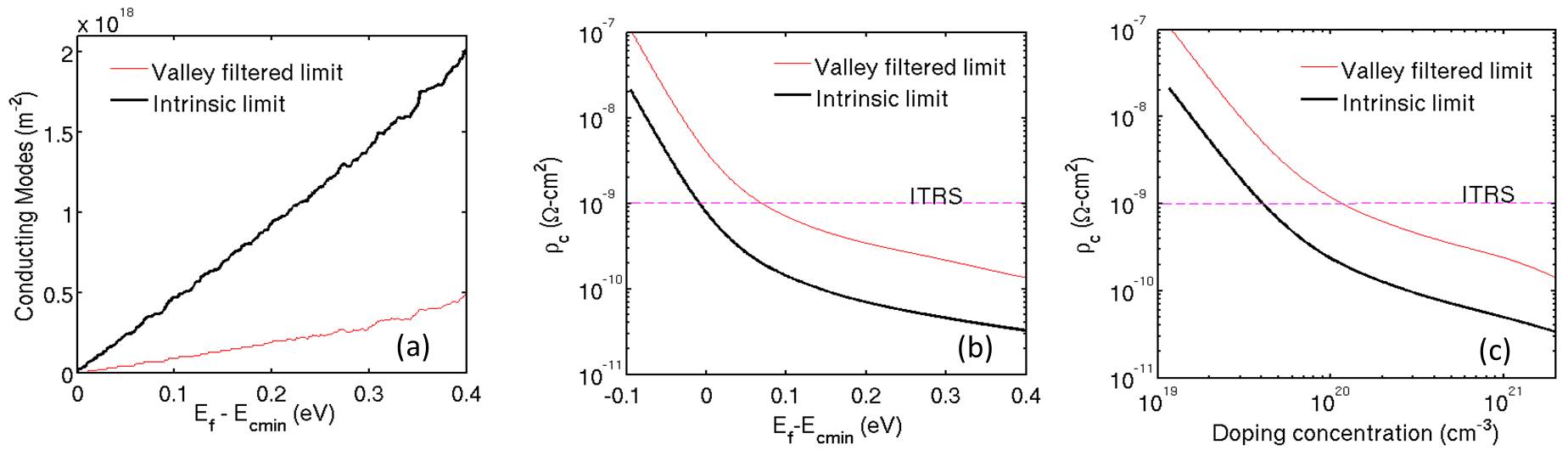}
\caption{(a) Density of available conducting modes in [100] oriented Si as a function of the offset of the Fermi level $E_f$ and the conduction band minimum $E_c$. (b) Lower limit of  contact resistivity $\rho_{c}$ as a function of the offset of the Fermi Level and conduction band minimum. (c) $\rho_c$ as a function of doping concentration in $n$-Si. It is evident from these figures that valley filtering raises the floor of the lowest contact resistivity that can be achieved in $n$-Si by at least an order of magnitude at high doping concentrations by severely reducing the number of modes in Si available for conduction due to mode-space mismatch with the metal.}
\label{fig:Fig3}
\end{figure*}

From this information, the lower limit of single-contact resistivity \cite{comment} ($\rho_{c}$) of Si is computed in the Landauer formalism. Using the notation of Maassen et al.\cite{maassen2013full} this is given by
\begin{equation}
\label{eq:Eq2}
\frac{1}{\rho_{c}}=\frac{2q^2}{h}\int_{-\infty}^{\infty} M(E)(\frac{-\partial f}{\partial E})dE   \quad(\Omega^{-1}m^{-2})		
\end{equation}
Here, $f$ is the Fermi-Dirac distribution function and $\frac{2q^2}{h}$ is the fundamental quantum of conductance. In $n$-Si, the occupancy of available conducting modes $M(E)$ is modulated by donor doping (and temperature, assumed to be 300K throughout), which changes the Fermi level $E_{f}$ and consequently changes $f$ and $\rho_{c}$. The precise dependence of the position of the Fermi level on doping concentration ($N_d$ cm$^{-3}$) is computed using the ATK package. The Si unit cell is populated by an additional negative charge corresponding to $N_d$ and compensated by a positive charge of the same amount to maintain neutrality. The Fermi Level is then computed by integrating the bulk band structure within the 1st BZ of the Si FCC unit cell using a uniform $k$-grid of 20$^3$ points until the desired charge corresponding to $N_d$ is reached.

A very important consideration in interface transmission problems is that of momentum conservation. For interface transmission considered in this letter, where the boundary conditions transverse to the direction of transport are periodic, transverse momenta on either side of the interface are conserved. The allowed transverse momenta, however, are \textit{not} the individual transverse momenta for the metal or semiconductor considered separately. Instead the allowed transverse momenta are determined by the \textit{non-primitive transverse interface unit cell vectors}\cite{boykin2005practical}. Consequently, the allowed transverse momenta for the [100] oriented CoSi$_{2}$-Si unit cells bound by the [010] and [001] directions in the TBZ are \textit{very different} from those for the unit cell bound by [0-11] and [011] directions. 

The importance of this feature for eventual interface electronic structure and transport cannot be overemphasized.  Since the allowed momenta in the TBZ are strongly dependent on choice of unit cell vector, so is the transmission spectrum. This is very evident from Figure \ref{fig:Fig1}, where the constant energy ellipsoids in bulk Si close to the bottom of the conduction band get projected along different transverse directions. More importantly, this effect manifests as a significant change in the projection of the Fermi surface of CoSi$_{2}$ on to the TBZ bound by the [010]/[001] and [0-11]/[011] orientations, respectively. The CoSi$_{2}$ transmission spectrum in the [010]/[001] case (Figure \ref{fig:Fig1}(a)) overlaps with its corresponding Si transmission spectrum completely. In other words, the 'ideal metal' assumption holds completely since every semiconducting mode is supplied by at least one mode from the metal. For the [0-11]/[011] case (Figure \ref{fig:Fig1}(b)), however, it is evident that the 'ideal metal' assumption fails. The Si ellipses at the four corners of the TBZ find no supplying mode from the metal. The term 'valley filtering' is therefore apt in this case and is a consequence of the metal being a non-ideal supplier of conducting modes into Si.

The effect of semiconducting valley-filtering on the transmission spectra of the coupled interfaces is shown in Figure \ref{fig:Fig2}. The only modes that conduct in the [0-11]/[011] bound coupled interface are the central modes that find supplying modes from the metal. In sharp contrast, the transmission spectrum of the [010]/[001] bound [100] coupled interface is identical to that in bulk Si. This effect is quantified in Figure \ref{fig:Fig3}. A significant reduction in the number of \textit{available} modes for conduction is seen due to the valley filtering effect. From equation \ref{eq:Eq2}, it is evident that a reduction in the number of available modes adversely impacts contact resistivity, raising the minimum achievable contact resistivity in Si significantly. For instance, from Figure \ref{fig:Fig3}(c) it can be seen that at $N_d=$ 7$\times$10$^{20}$cm$^{-3}$, the minimum achievable $\rho_{c}$ in $n$-Si increases a full order of magnitude from 6$\times$10$^{-11}$$\Omega$-cm$^{2}$ to about 3$\times$10$^{-10}$$\Omega$-cm$^{2}$ due to valley filtering in the [0-11]/[011] bound CoSi$_{2}$-Si interface. 

It must be reiterated that the calculated $\rho_c$ limits are reported assuming ballistic transmission across the interface. The presence of potential barriers of finite height and width, lattice imperfections and back-scattering at the interface due to effective mass mismatch between the metal and semiconductor will attenuate the transmission and consequently increase the contact resistivity considerably. Thus, while the ITRS requirement remains attainable at high doping concentrations, valley-filtering (if it does exist at a specific interface) in combination with the above limiting factors may prove to be a fundamental obstacle in improving contact resistivity.

It is important to recognize the limits of applicability of such an analysis. The electronic structure of metals changes considerably with confinement and strain \cite{CuConfinement, CuModel}. This may affect valley filtering considerably. Also, the choice of simulation cell in this letter was guided by experiment. The simulation cells were chosen as the smallest lattice-matched unit cells obeying the 2D periodicity conditions indicated by the experiment. It is possible that other metal-semiconductor combinations have larger fundamental lattice-matched unit cells and different TBZ overlaps. For M-S combinations where experimental data is not available, the only alternative may be simulating interface formation through molecular dynamics (MD) simulations and then computing electronic transport on the resultant structures.

%Valley-filtering is dependent on non-overlapping transverse momenta. In most analyses of metal-semiconductor transport problems, 2D-periodicity is assumed transverse to the direction of transport. Depending on the contact geometry, this assumption may or may not hold. For instance, the valley filtering effect will not manifest if a contact confined quantum mechanically in two dimensions is simulated. .  

In conclusion, the lower limits of resistivity are computed in $n$-Si for CoSi$_{2}$-Si interface structures. It is shown that in some situations, the metal may behave 'non-ideally' i.e. not supply conducting modes to the semiconductor equally throughout the TBZ. This effect is shown to be dependent on the specific interface chemistry simulated. The filtering of semiconducting valleys is shown to increase the minimum achievable contact resistivity in Si by an order of magnitude at high doping concentrations. While emphasizing the need to go beyond semi-classical and effective mass based electronic transport, the results highlight the need to revise the existing atomistic device simulation paradigm \cite{fonseca2013efficient}, where metals are not explicitly included in the simulation. Including metals explicitly in atomistic DFT or TB simulations will ensure that effects such as valley filtering are incorporated into contact/device simulations.
 
From a technology perspective, these results indicate that in addition to metrics commonly understood to affect contact resistivity - SBH, barrier width, doping concentration, contact area - it is worth carefully considering if valley-filtering plays a role at the interface of a candidate metal-semiconductor interface. 

%It can be seen that the The first is related to the effect of the interface chemistry being simulated. It is very evident that the transmission spectra for [100] oriented bulk CoSi$_{2}$ and Si shown in figure \ref{fig:Fig1} change quite significantly depending on the boundary conditions of the TBZ. 
%This seemingly unintuitive result is explained by appealing to the concept of Brillouin Zone Folding (BZF) \cite{boykin2005practical}. BZF enables the determination of the allowed $k$-vectors in a non-primitive unit cell for a given crystal structure. Once these are calculated, the band structure of a primitive unit cell is then projected onto the non-primitive BZ. In [100] oriented Si bound by [010] and [001] orientations, the six constant energy ellipsoids in Si get projected onto the [100] unit cell vectors. 2 of these are in the direction of transport i.e. [100] and . The FS at the conduction band minimum in bulk Si, which is a collection of six degenerate ellipsoids along the $X$ orientation gets projected onto non-primitive unit cell vectors that form the TBZ for the non-primitive unit-cell. Since the TBZ is bound by [100]-type orientations in one case and [011]-type directions in the other, the allowed non-

The authors thank Jorge Kittl, Borna Obradovic, Ryan Hatcher and Mark Rodder for helpful discussions.

\nocite{*}
\bibliography{bibliography}% Produces the bibliography via BibTeX.

\begin{thebibliography}{18}%
\makeatletter
\providecommand \@ifxundefined [1]{%
 \@ifx{#1\undefined}
}%
\providecommand \@ifnum [1]{%
 \ifnum #1\expandafter \@firstoftwo
 \else \expandafter \@secondoftwo
 \fi
}%
\providecommand \@ifx [1]{%
 \ifx #1\expandafter \@firstoftwo
 \else \expandafter \@secondoftwo
 \fi
}%
\providecommand \natexlab [1]{#1}%
\providecommand \enquote  [1]{``#1''}%
\providecommand \bibnamefont  [1]{#1}%
\providecommand \bibfnamefont [1]{#1}%
\providecommand \citenamefont [1]{#1}%
\providecommand \href@noop [0]{\@secondoftwo}%
\providecommand \href [0]{\begingroup \@sanitize@url \@href}%
\providecommand \@href[1]{\@@startlink{#1}\@@href}%
\providecommand \@@href[1]{\endgroup#1\@@endlink}%
\providecommand \@sanitize@url [0]{\catcode `\\12\catcode `\$12\catcode
  `\&12\catcode `\#12\catcode `\^12\catcode `\_12\catcode `\%12\relax}%
\providecommand \@@startlink[1]{}%
\providecommand \@@endlink[0]{}%
\providecommand \url  [0]{\begingroup\@sanitize@url \@url }%
\providecommand \@url [1]{\endgroup\@href {#1}{\urlprefix }}%
\providecommand \urlprefix  [0]{URL }%
\providecommand \Eprint [0]{\href }%
\providecommand \doibase [0]{http://dx.doi.org/}%
\providecommand \selectlanguage [0]{\@gobble}%
\providecommand \bibinfo  [0]{\@secondoftwo}%
\providecommand \bibfield  [0]{\@secondoftwo}%
\providecommand \translation [1]{[#1]}%
\providecommand \BibitemOpen [0]{}%
\providecommand \bibitemStop [0]{}%
\providecommand \bibitemNoStop [0]{.\EOS\space}%
\providecommand \EOS [0]{\spacefactor3000\relax}%
\providecommand \BibitemShut  [1]{\csname bibitem#1\endcsname}%
\let\auto@bib@innerbib\@empty
%</preamble>
\bibitem [{itr(2011)}]{itrs}%
  \BibitemOpen
  \href {http://www.itrs.net/} {\enquote {\bibinfo {title} {International
  technology roadmap for semiconductors - front end processes},}\ } (\bibinfo
  {year} {2011})\BibitemShut {NoStop}%
\bibitem [{\citenamefont {Yu}(1970)}]{yu1970electron}%
  \BibitemOpen
  \bibfield  {author} {\bibinfo {author} {\bibfnamefont {A.}~\bibnamefont
  {Yu}},\ }\bibfield  {title} {\enquote {\bibinfo {title} {Electron tunneling
  and contact resistance of metal-silicon contact barriers},}\ }\href@noop {}
  {\bibfield  {journal} {\bibinfo  {journal} {Solid-State Electronics}\
  }\textbf {\bibinfo {volume} {13}},\ \bibinfo {pages} {239--247} (\bibinfo
  {year} {1970})}\BibitemShut {NoStop}%
\bibitem [{\citenamefont {Griffiths}(2013)}]{griffiths2013introduction}%
  \BibitemOpen
  \bibfield  {author} {\bibinfo {author} {\bibfnamefont {D.~J.}\ \bibnamefont
  {Griffiths}},\ }\href@noop {} {\emph {\bibinfo {title} {Introduction to
  quantum mechanics: pearson new international edition}}}\ (\bibinfo
  {publisher} {Pearson Education Limited},\ \bibinfo {year} {2013})\BibitemShut
  {NoStop}%
\bibitem [{\citenamefont {Maassen}\ \emph {et~al.}(2013)\citenamefont
  {Maassen}, \citenamefont {Jeong}, \citenamefont {Baraskar}, \citenamefont
  {Rodwell},\ and\ \citenamefont {Lundstrom}}]{maassen2013full}%
  \BibitemOpen
  \bibfield  {author} {\bibinfo {author} {\bibfnamefont {J.}~\bibnamefont
  {Maassen}}, \bibinfo {author} {\bibfnamefont {C.}~\bibnamefont {Jeong}},
  \bibinfo {author} {\bibfnamefont {A.}~\bibnamefont {Baraskar}}, \bibinfo
  {author} {\bibfnamefont {M.}~\bibnamefont {Rodwell}}, \ and\ \bibinfo
  {author} {\bibfnamefont {M.}~\bibnamefont {Lundstrom}},\ }\bibfield  {title}
  {\enquote {\bibinfo {title} {Full band calculations of the intrinsic lower
  limit of contact resistivity},}\ }\href@noop {} {\bibfield  {journal}
  {\bibinfo  {journal} {Applied Physics Letters}\ }\textbf {\bibinfo {volume}
  {102}},\ \bibinfo {pages} {111605--111605} (\bibinfo {year}
  {2013})}\BibitemShut {NoStop}%
\bibitem [{\citenamefont {Trumbore}(1960)}]{trumbore1960solid}%
  \BibitemOpen
  \bibfield  {author} {\bibinfo {author} {\bibfnamefont {F.~A.}\ \bibnamefont
  {Trumbore}},\ }\bibfield  {title} {\enquote {\bibinfo {title} {Solid
  solubilities of impurity elements in germanium and silicon*},}\ }\href@noop
  {} {\bibfield  {journal} {\bibinfo  {journal} {Bell System Technical
  Journal}\ }\textbf {\bibinfo {volume} {39}},\ \bibinfo {pages} {205--233}
  (\bibinfo {year} {1960})}\BibitemShut {NoStop}%
\bibitem [{\citenamefont {Weber}(2013)}]{weber2013importance}%
  \BibitemOpen
  \bibfield  {author} {\bibinfo {author} {\bibfnamefont {C.}~\bibnamefont
  {Weber}},\ }\bibfield  {title} {\enquote {\bibinfo {title} {The importance of
  metal transverse momentum for silicon contact resistivity},}\ }\href@noop {}
  {\bibfield  {journal} {\bibinfo  {journal} {Applied Physics Letters}\
  }\textbf {\bibinfo {volume} {103}},\ \bibinfo {pages} {193505} (\bibinfo
  {year} {2013})}\BibitemShut {NoStop}%
\bibitem [{\citenamefont {Park}\ \emph {et~al.}(2013)\citenamefont {Park},
  \citenamefont {Kharche}, \citenamefont {Basu}, \citenamefont {Jiang},
  \citenamefont {Nayak}, \citenamefont {Weber}, \citenamefont {Hegde},
  \citenamefont {Haume}, \citenamefont {Kubis}, \citenamefont {Povolotskyi}
  \emph {et~al.}}]{park2013scaling}%
  \BibitemOpen
  \bibfield  {author} {\bibinfo {author} {\bibfnamefont {S.-H.}\ \bibnamefont
  {Park}}, \bibinfo {author} {\bibfnamefont {N.}~\bibnamefont {Kharche}},
  \bibinfo {author} {\bibfnamefont {D.}~\bibnamefont {Basu}}, \bibinfo {author}
  {\bibfnamefont {Z.}~\bibnamefont {Jiang}}, \bibinfo {author} {\bibfnamefont
  {S.}~\bibnamefont {Nayak}}, \bibinfo {author} {\bibfnamefont
  {C.}~\bibnamefont {Weber}}, \bibinfo {author} {\bibfnamefont
  {G.}~\bibnamefont {Hegde}}, \bibinfo {author} {\bibfnamefont
  {K.}~\bibnamefont {Haume}}, \bibinfo {author} {\bibfnamefont
  {T.}~\bibnamefont {Kubis}}, \bibinfo {author} {\bibfnamefont
  {M.}~\bibnamefont {Povolotskyi}},  \emph {et~al.},\ }\bibfield  {title}
  {\enquote {\bibinfo {title} {Scaling effect on specific contact resistivity
  in nano-scale metal-semiconductor contacts},}\ }in\ \href@noop {} {\emph
  {\bibinfo {booktitle} {Device Research Conference (DRC), 2013 71st Annual}}}\
  (\bibinfo {organization} {IEEE},\ \bibinfo {year} {2013})\ pp.\ \bibinfo
  {pages} {125--126}\BibitemShut {NoStop}%
\bibitem [{\citenamefont {Gao}\ and\ \citenamefont {Guo}(2012)}]{gao2012ab}%
  \BibitemOpen
  \bibfield  {author} {\bibinfo {author} {\bibfnamefont {Q.}~\bibnamefont
  {Gao}}\ and\ \bibinfo {author} {\bibfnamefont {J.}~\bibnamefont {Guo}},\
  }\bibfield  {title} {\enquote {\bibinfo {title} {Ab initio quantum transport
  simulation of silicide-silicon contacts},}\ }\href@noop {} {\bibfield
  {journal} {\bibinfo  {journal} {Journal of Applied Physics}\ }\textbf
  {\bibinfo {volume} {111}},\ \bibinfo {pages} {014305} (\bibinfo {year}
  {2012})}\BibitemShut {NoStop}%
\bibitem [{\citenamefont {Tung}, \citenamefont {Schrey},\ and\ \citenamefont
  {Yalisove}(1989)}]{tung1989homoepitaxial}%
  \BibitemOpen
  \bibfield  {author} {\bibinfo {author} {\bibfnamefont {R.}~\bibnamefont
  {Tung}}, \bibinfo {author} {\bibfnamefont {F.}~\bibnamefont {Schrey}}, \ and\
  \bibinfo {author} {\bibfnamefont {S.}~\bibnamefont {Yalisove}},\ }\bibfield
  {title} {\enquote {\bibinfo {title} {Homoepitaxial growth of cosi2 and nisi2
  on (100) and (110) surfaces at room temperature},}\ }\href@noop {} {\bibfield
   {journal} {\bibinfo  {journal} {Applied Physics Letters}\ }\textbf {\bibinfo
  {volume} {55}},\ \bibinfo {pages} {2005--2007} (\bibinfo {year}
  {1989})}\BibitemShut {NoStop}%
\bibitem [{\citenamefont {Boykin}, \citenamefont {Klimeck},\ and\ \citenamefont
  {Oyafuso}(2004)}]{boykin2004valence}%
  \BibitemOpen
  \bibfield  {author} {\bibinfo {author} {\bibfnamefont {T.~B.}\ \bibnamefont
  {Boykin}}, \bibinfo {author} {\bibfnamefont {G.}~\bibnamefont {Klimeck}}, \
  and\ \bibinfo {author} {\bibfnamefont {F.}~\bibnamefont {Oyafuso}},\
  }\bibfield  {title} {\enquote {\bibinfo {title} {Valence band effective-mass
  expressions in the sp 3 d 5 s* empirical tight-binding model applied to a si
  and ge parametrization},}\ }\href@noop {} {\bibfield  {journal} {\bibinfo
  {journal} {Physical Review B}\ }\textbf {\bibinfo {volume} {69}},\ \bibinfo
  {pages} {115201} (\bibinfo {year} {2004})}\BibitemShut {NoStop}%
\bibitem [{\citenamefont {Perdew}, \citenamefont {Burke},\ and\ \citenamefont
  {Ernzerhof}(1996)}]{perdew1996generalized}%
  \BibitemOpen
  \bibfield  {author} {\bibinfo {author} {\bibfnamefont {J.~P.}\ \bibnamefont
  {Perdew}}, \bibinfo {author} {\bibfnamefont {K.}~\bibnamefont {Burke}}, \
  and\ \bibinfo {author} {\bibfnamefont {M.}~\bibnamefont {Ernzerhof}},\
  }\bibfield  {title} {\enquote {\bibinfo {title} {Generalized gradient
  approximation made simple},}\ }\href@noop {} {\bibfield  {journal} {\bibinfo
  {journal} {Physical review letters}\ }\textbf {\bibinfo {volume} {77}},\
  \bibinfo {pages} {3865} (\bibinfo {year} {1996})}\BibitemShut {NoStop}%
\bibitem [{atk(2014)}]{atk}%
  \BibitemOpen
  \href {http://www.quantumwise.com} {\enquote {\bibinfo {title} {Atomistix
  toolkit version 13.8.0 quantumwise a/s quantumwise},}\ } (\bibinfo {year}
  {2014})\BibitemShut {NoStop}%
\bibitem [{com()}]{comment}%
  \BibitemOpen
  \bibfield  {title} {\enquote {\bibinfo {title} {The itrs requirement for
  contact resistivity is for a single contact although contact resistivity
  measurements contain the effect of source and drain contact resistances},}\
  }\href@noop {} {\ }\BibitemShut {NoStop}%
\bibitem [{\citenamefont {Boykin}\ and\ \citenamefont
  {Klimeck}(2005)}]{boykin2005practical}%
  \BibitemOpen
  \bibfield  {author} {\bibinfo {author} {\bibfnamefont {T.~B.}\ \bibnamefont
  {Boykin}}\ and\ \bibinfo {author} {\bibfnamefont {G.}~\bibnamefont
  {Klimeck}},\ }\bibfield  {title} {\enquote {\bibinfo {title} {Practical
  application of zone-folding concepts in tight-binding calculations},}\
  }\href@noop {} {\bibfield  {journal} {\bibinfo  {journal} {Physical Review
  B}\ }\textbf {\bibinfo {volume} {71}},\ \bibinfo {pages} {115215} (\bibinfo
  {year} {2005})}\BibitemShut {NoStop}%
\bibitem [{\citenamefont {Hegde}\ \emph
  {et~al.}(2014{\natexlab{a}})\citenamefont {Hegde}, \citenamefont
  {Povolotskyi}, \citenamefont {Kubis}, \citenamefont {Charles},\ and\
  \citenamefont {Klimeck}}]{CuConfinement}%
  \BibitemOpen
  \bibfield  {author} {\bibinfo {author} {\bibfnamefont {G.}~\bibnamefont
  {Hegde}}, \bibinfo {author} {\bibfnamefont {M.}~\bibnamefont {Povolotskyi}},
  \bibinfo {author} {\bibfnamefont {T.}~\bibnamefont {Kubis}}, \bibinfo
  {author} {\bibfnamefont {J.}~\bibnamefont {Charles}}, \ and\ \bibinfo
  {author} {\bibfnamefont {G.}~\bibnamefont {Klimeck}},\ }\bibfield  {title}
  {\enquote {\bibinfo {title} {An environment-dependent semi-empirical tight
  binding model suitable for electron transport in bulk metals, metal alloys,
  metallic interfaces, and metallic nanostructures. ii. application—effect of
  quantum confinement and homogeneous strain on cu conductance},}\ }\href
  {\doibase http://dx.doi.org/10.1063/1.4868979} {\bibfield  {journal}
  {\bibinfo  {journal} {Journal of Applied Physics}\ }\textbf {\bibinfo
  {volume} {115}},\ \bibinfo {eid} {123704} (\bibinfo {year}
  {2014}{\natexlab{a}})}\BibitemShut {NoStop}%
\bibitem [{\citenamefont {Hegde}\ \emph
  {et~al.}(2014{\natexlab{b}})\citenamefont {Hegde}, \citenamefont
  {Povolotskyi}, \citenamefont {Kubis}, \citenamefont {Boykin},\ and\
  \citenamefont {Klimeck}}]{CuModel}%
  \BibitemOpen
  \bibfield  {author} {\bibinfo {author} {\bibfnamefont {G.}~\bibnamefont
  {Hegde}}, \bibinfo {author} {\bibfnamefont {M.}~\bibnamefont {Povolotskyi}},
  \bibinfo {author} {\bibfnamefont {T.}~\bibnamefont {Kubis}}, \bibinfo
  {author} {\bibfnamefont {T.}~\bibnamefont {Boykin}}, \ and\ \bibinfo {author}
  {\bibfnamefont {G.}~\bibnamefont {Klimeck}},\ }\bibfield  {title} {\enquote
  {\bibinfo {title} {An environment-dependent semi-empirical tight binding
  model suitable for electron transport in bulk metals, metal alloys, metallic
  interfaces, and metallic nanostructures. i. model and validation},}\ }\href
  {\doibase http://dx.doi.org/10.1063/1.4868977} {\bibfield  {journal}
  {\bibinfo  {journal} {Journal of Applied Physics}\ }\textbf {\bibinfo
  {volume} {115}},\ \bibinfo {eid} {123703} (\bibinfo {year}
  {2014}{\natexlab{b}})}\BibitemShut {NoStop}%
\bibitem [{\citenamefont {Fonseca}\ \emph {et~al.}(2013)\citenamefont
  {Fonseca}, \citenamefont {Kubis}, \citenamefont {Povolotskyi}, \citenamefont
  {Novakovic}, \citenamefont {Ajoy}, \citenamefont {Hegde}, \citenamefont
  {Ilatikhameneh}, \citenamefont {Jiang}, \citenamefont {Sengupta},
  \citenamefont {Tan} \emph {et~al.}}]{fonseca2013efficient}%
  \BibitemOpen
  \bibfield  {author} {\bibinfo {author} {\bibfnamefont {J.}~\bibnamefont
  {Fonseca}}, \bibinfo {author} {\bibfnamefont {T.}~\bibnamefont {Kubis}},
  \bibinfo {author} {\bibfnamefont {M.}~\bibnamefont {Povolotskyi}}, \bibinfo
  {author} {\bibfnamefont {B.}~\bibnamefont {Novakovic}}, \bibinfo {author}
  {\bibfnamefont {A.}~\bibnamefont {Ajoy}}, \bibinfo {author} {\bibfnamefont
  {G.}~\bibnamefont {Hegde}}, \bibinfo {author} {\bibfnamefont
  {H.}~\bibnamefont {Ilatikhameneh}}, \bibinfo {author} {\bibfnamefont
  {Z.}~\bibnamefont {Jiang}}, \bibinfo {author} {\bibfnamefont
  {P.}~\bibnamefont {Sengupta}}, \bibinfo {author} {\bibfnamefont
  {Y.}~\bibnamefont {Tan}},  \emph {et~al.},\ }\bibfield  {title} {\enquote
  {\bibinfo {title} {Efficient and realistic device modeling from atomic detail
  to the nanoscale},}\ }\href@noop {} {\bibfield  {journal} {\bibinfo
  {journal} {Journal of Computational Electronics}\ }\textbf {\bibinfo {volume}
  {12}},\ \bibinfo {pages} {592--600} (\bibinfo {year} {2013})}\BibitemShut
  {NoStop}%
\bibitem [{\citenamefont {Datta}(1997)}]{datta1997electronic}%
  \BibitemOpen
  \bibfield  {author} {\bibinfo {author} {\bibfnamefont {S.}~\bibnamefont
  {Datta}},\ }\href@noop {} {\emph {\bibinfo {title} {Electronic transport in
  mesoscopic systems}}}\ (\bibinfo  {publisher} {Cambridge university press},\
  \bibinfo {year} {1997})\BibitemShut {NoStop}%
\end{thebibliography}%

\end{document}